# Providing information can be a stable non-cooperative evolutionary strategy

Jean-Louis Dessalles

Telecom ParisTech
46 rue Barrault – F-75013 Paris, France
dessalles@telecom-paristech.fr - www.telecom-paristech.fr/~jld

### [**Abstract**]

Human language is still an embarrassment for evolutionary theory, as the speaker's benefit remains unclear. The willingness to communicate information is shown here to be an evolutionary stable strategy (ESS), even if acquiring original information from the environment involves significant cost and communicating it provides no material benefit to addressees. In this study, communication is used to *advertise* the emitter's ability to obtain novel information. We found that communication strategies can take two forms, competitive and uniform, that these two strategies are stable and that they necessarily coexist.

## Donner des informations peut être une stratégie non-cooperative évolutionnairement stable

Jean-Louis Dessalles

Telecom ParisTech
46 rue Barrault – F-75013 Paris, France
dessalles@telecom-paristech.fr - www.telecom-paristech.fr/~jld

### [**Résumé**]

Le langage humain demeure une énigme dans le cadre de la théorie de l'évolution, car le bénéfice du locuteur demeure incertain. Je montre que la propension à communiquer des informations est une stratégie évolutionnairement stable (ESS), même si acquérir des informations originales comporte un coût et même si ces informations ne procurent aucun bénéfice matériel à ceux à qui elles sont communiquées. Dans cette étude, la communication est utilisée pour afficher la capacité du locuteur à obtenir des informations nouvelles. Il se trouve que les stratégies de communication sont au nombre de deux : compétitive et uniforme, que ces deux stratégies sont stables et qu'elles coexistent nécessairement.

# Introduction

Providing information to conspecifics is a distinctive feature of the human species. It is mainly observed in spontaneous conversation, which a massive [*1*] and universal [*2*] behaviour, but also currently in blogging and micro-blogging platforms [*3*], on technical online forums and, in a more elaborate form, in open source software communities [*4*]. Providing information is a costly behaviour: it requires time and efforts to get original information worth to tell, and communicating it through language takes up a considerable amount of available time [*1*]. Giving potentially useful information to conspecifics is therefore a form of altruism. A challenge for evolutionary biology is to provide a logical and mathematical account showing that communicating information can be an evolutionary stable strategy (ESS).

# Human Language and altruism

There are many known cases of altruism in nature, but language resembles none of them. Several facts about language behavior challenge traditional theories of biological altruism (Table 1):

O1. Individuals devote 30% of their awake time to language [*1*] in various cultures [*2*], and speak some 15 000 words each day on average [*5*].

O2. The price of information is low, due to the presence of talkative individuals [*6*].

O3. Many accepted utterances are about futile matters that are unlikely to have impact on listeners' lives.

O4. Listeners are not (or loosely) discriminated by speakers; speech is directed towards several individuals simultaneously [*7*].

O5. A significant share of talking time is devoted to signaling immediate or past unexpected events [*8–10*]; Drawing conspecifics' attention toward unlikely situations is a distinctive behavior of our species [*11, 12*] that shows up early in ontogeny [*13*].

O6. Human beings learn and understand tens of thousands of words and set phrases.

O7. Social bonds are highly correlated with sharing conversational time [*14*].

O8. There is no major language difference depending on sex [*5*].

O9. Other primates do not systematically share information about events.

**Table 1.** Several theories of costly altruistic (or apparently altruistic) behavior are compared by their ability to account for the nine observations O1-9 about human language. A plus (+) sign means that the theory correctly predicts the observation, a 0 means that it does not predict it, a minus sign means that it conflicts with it.

|  |  | O1 | O2 | O3 | O4 | O5 | O6 | O7 | O8 | O9 |
|---|---|---|---|---|---|---|---|---|---|---|
| cooperative | Group selection [*15*] | − | 0 | − | + | 0 | 0 | 0 | + | 0 |
| | Parochial altruism [*16*] | − | 0 | − | + | 0 | 0 | 0 | − | 0 |
| | Reciprocal Cooperation [*17*] | 0 | − | − | − | 0 | 0 | + | + | 0 |
| | Indirect reciprocity [*17*] | 0 | − | − | − | 0 | 0 | 0 | + | 0 |
| | Network reciprocity [*17*] | 0 | − | − | − | 0 | 0 | + | + | 0 |
| non-cooperative | Kin selection [*18*] | 0 | 0 | − | − | 0 | 0 | 0 | − | 0 |
| | Sexual selection [*19*] | + | + | + | + | 0 | + | 0 | − | 0 |
| | Costly signaling [*20*] | − | + | 0 | + | 0 | 0 | + | + | 0 |

Group selection requires significant discrepancies between groups to operate [*15*], what fact O1 contradicts. Its parochial version [*16*], reserved to warriors, also conflicts with the fact that language is equally shared among genders (O8). Moreover, these group-selectionist models, as well as all cooperative scenarios, require that the altruistic behavior provide substantial benefit to recipients [*17*]. Language does not match this expectation (fact O3). Cooperative scenarios are also at odds with facts O2 and O4, as they crucially rely on the detection of free-riders who take information and fail to return the equivalent of what they received. If information has a price, it should not be given for free without discrimination. Yet, language is more like broadcast rather than like whispering [*7*]. Information is offered more often than demanded. "People compete to say things. They strive to be heard. [...] Those who fail to yield the floor to their colleagues are considered selfish, not altruistic." [*19*] Kin selection would also demand efficiency and discrimination. In addition, if language evolved to talk to offspring [*18*], then (O8) is unexpected as parental investment varies depending on sex. (O8) conflicts with the sexual selection model as well. Lastly, all theories listed in Table 1 need external hypotheses to explain language uniqueness (O9).

Modified versions of these classical theories of altruism might possibly better explain how the urge to communicate information to conspecifics could evolve. We found such a consistent account within the costly signaling framework.

# Human Language as Costly Signal

Costly signaling theory (CST) [*20*, *21*] describes signals as competitive. In its social version [*20*, *22*], individuals display some quality $q$ in order to be chosen as coalition partners. In bird species such as *turdoides squamiceps*, $q$ may be the ability to mob predators [*21*]. Signalers benefit from being joined or accepted by 'followers', whereas followers benefit from joining individuals with high $q$. Such signaling system may evolve to a stable honest state, where signal intensity is an increasing function of $q$ [*20*].

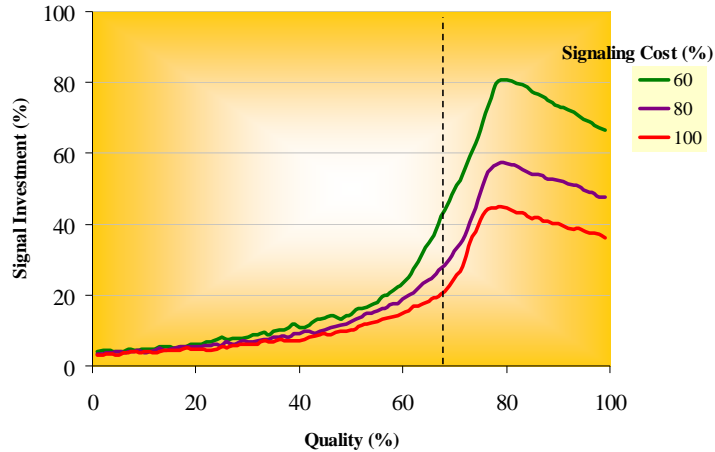

**Figure 1.** Simulation of a social version of CST.
Individuals differ by their investment in signaling (y-axis) depending on their quality (x-axis). The three curves differ by the cost coefficient $C$.

Social-CST may explain several important aspects of language behavior (Table 1), but as it stands, it wrongly predicts that only a minority of individuals will talk, in contradiction with (O1). Signaling competition generates a threshold in quality, below which it is not worth while investing in signaling (Figure 1). Each individual of quality $q \in [0,1]$ emits a signal $s(q) = g(q)\,q$, where $g(q) \in [0,1]$ is the willingness to signal. Signaling demands proportional cost $c = C\,g(q)$. We can see on Figure 1 that low quality individuals are discouraged from investing in signaling. The signaling threshold, when quality is uniformly distributed, is $\eta = 1 - 1/n$, where $n$ is the maximal number of recruited individuals per signaler. Figures 1 and 2A have been obtained by artificially limiting n to 3, otherwise all individuals in the population would have ended up following only two or three top signalers. Figure 2A illustrates the 'star-system' generated by CST, as high-quality individuals attract most followers, while low-quality individuals have no interest in wasting energy in hopeless signaling. Note that $\eta$ does not depend on cost coefficient $C$.

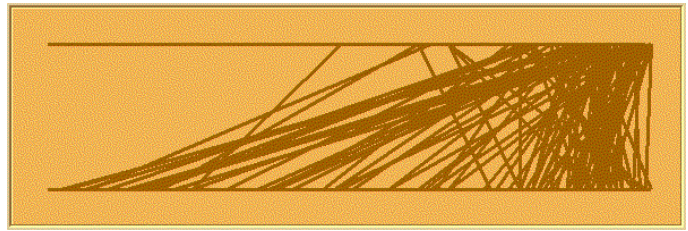

(A)

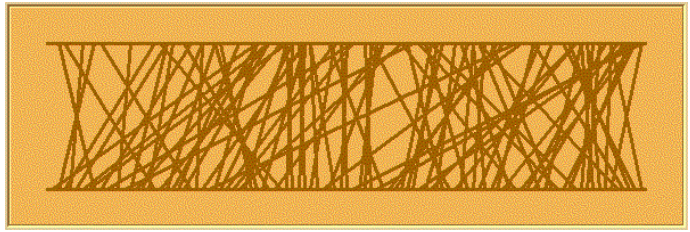

(B)

**Figure 2.** Examples of sociograms for classical social-CST (A) and the FSM model (B). Individuals in the population are located twice, on lower and upper horizontal axes, depending on their quality. Social links are represented by lines between the lower and the upper axis (only links to best friends are shown).

## Competitive Friendship Signaling

We could develop a version of CST in which all individuals, even low-quality ones, do signal, and benefit from it (Figure 3A). The solution we adopted was inspired by Dunbar's comparison of language with grooming [*14*]. The model, which can be dubbed Friendship Signaling Model (FSM), relies on two additional hypotheses. First, individuals symmetrically appraise each other's signaling performance before deciding to become friends. Second, social bonds are constrained by the amount of time individuals can offer to their friends. Individuals therefore base their decision to join each other not on the sole signal they emit, but also on the time they are ready to share.

When two individuals A and B meet, they negotiate a rank $i$ in their friendship hierarchy, based on the other's "social offer." A's social offer to B amounts to $s(q_A) \, r^i$, where $s(q_A)$ is the signal displayed by A (note that $q_A$ is unknown to B), $i$ is B's potential rank in A's friendship list and $r^i$ is the amount of time offered to B ($0 \leq r \leq 1$).

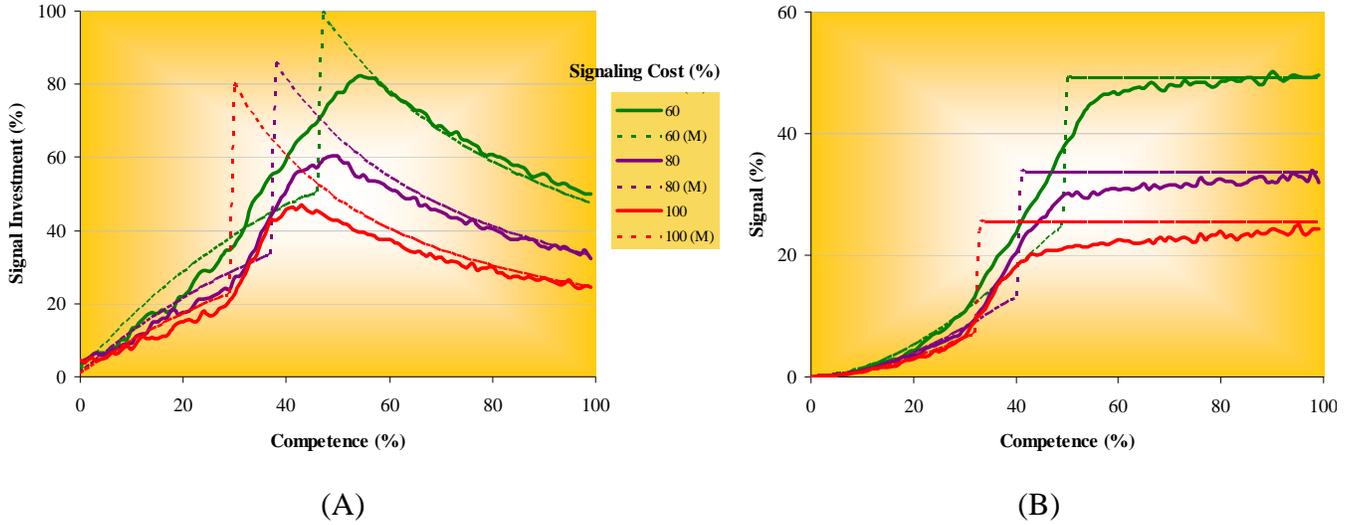

(A) (B)

**Figure 3.** Investment in communication (A) and signals (B) in the friendship scenario. Dashed lines show model predictions. The model does not include the transition between the two ESS (see annex). The number of friends per individual is limited to 3, for comparison with Figure 1.

Due to this two-way negotiation, competition on the social relationship "market" rapidly leads to assortative links, as shown Figure 2B (see annex). In the assortative mode, individuals tend to bind to partners with similar social offer and therefore, at equilibrium, with similar competence. This situation corresponds to the increasing part of the curves in Figure 3. In that competence range, individuals publicly distinguish from each other by their competence. The benefit of this type of assortative bonding for an individual of quality $q$ is:

$$B_c(q) = P(q) - C \, s(q)/q$$

where $P(q)$ is the profit brought by an alliance with a partner of similar competence $q$. To be an ESS, this competitive state must be robust to unilateral change. Suppose a mutant with competence $q$ sends the signal normally sent with competence $q+dq$. The recruitment of a better partner provides $P(q+dq)$, by dint of an augmented cost $C \, s(q+dq)/q$. The benefit variation $dB_c = P'(q) \, dq - C \, s'(q) \, dq \, /q$ must be zero for the equilibrium to be stable, which gives: $s'(q) = q \, P'(q) \, / \, C$. In this competitive ESS, individuals emit:

$$s(q) = [ \, qP(q) - \int P(q) \, dq \, ] \, /C$$

If $P'(q) > 0$, *i.e.* if competence $q$ is relevant, there is no threshold for signaling (in contrast with standard CST): all individuals, including those in the lower competence range, benefit from communicating.

## Uniform Friendship Signaling

An unexpected finding of this study has been that competitive signaling coexists with another ESS. Individuals in the upper competence range ($q > \eta$) all send the same signal $s_m$ for a given cost coefficient (Figure 3B). In contrast to the competitive case, social bonding is no longer assortative, as individuals in the 'elite club' ($\eta < q \leq 1$) cannot distinguish from each other. Their profit is therefore $P((1+\eta)/2)$ on average. This uniform signaling mode is, understandably, an ESS. Those who unilaterally signal above $s_m$ pay an additional cost with no profit, since their partner will be a random member of the elite club anyway. Those who signal slightly below $s_m$ make a slight economy but their profit $P(\eta)$ is dramatically smaller than the normally expected $P((1+\eta)/2)$.

In this ideal situation, there is no way to determine the couple ($\eta$, $s_m$). All values of $s_m$ theoretically lead to an ESS. In the presence of noise, however, the indeterminacy collapses. $s_m$ gets a definite value (Figure 3B), at which the varying cost of signaling slightly above or below $s_m$ exactly compensates the varying risk of getting $P(\eta)$ instead of $P((1+\eta)/2)$ (see annex). For the profit function adopted in this study (see below), the value of the uniform signal $s_m$ is a decreasing function of the cost coefficient (Figure 3B), of noise amplitude and of the friend inequality factor $r$.

## The 'First To Tell' behavior and the origin of language

Though the FSM model could be applied to a variety of situations, including primate grooming, its main interest is to show how human language, or at least its potential precursors [26], can be an ESS. What needs to be shown is that the informational competence of the recruited friends and the social time they offer both increase survival. One known fact about hominin ecology may explain the two aspects.

At some point in hominin phylogeny, individuals became able to use weapons to kill at no risk [23–25]. The possibility of surprise killing dramatically disrupted traditional primate politics, based on physical dominance. In this new and unique context, the ability to anticipate surprise became an asset. By repeatedly demonstrating their informational competence through language, even about futile matters, individuals show off their ability to spot unexpected situations before others. In the CST framework, they do so to be recruited as social partners.

The shared time constraint ($r < 1$) is a natural consequence from the fact that information-competent friends provide protection. When two individuals become acquainted, each of them receives protection from the mere presence of the other. To represent this fact, $P$ can be given the following definition, which has been used in our simulations:

$$P(q) = 1 - \Pi_i (1 - K r^i q)$$

$K$ is a constant and the product is computed over all the individual's friends. This expression means that the presence of the $i^{th}$ friend during a fraction $r^i$ of the time contributes by $K r^i q$ to reducing the probability of getting killed.

This hypothesis concerning the role of information competence in the protection against surprise killing satisfies all the hypotheses of the FSM model. It also predicts the nine observations O1-O9 about language listed in Table 1. In comparison with standard CST, it explains why virtually all individuals talk (O1), why most utterances are inconsequential (O3), and why language is uniquely human (O9). The need to signal any form of unexpectedness explains in part why lexicon must be large and learned (O6). And most importantly, it explains why individuals feel the urge to be the first to tell about any unexpected event (O5).

This first-to-tell behavior, characteristic of many human conversations, suggests that language may have originated as a way to secure protection. It is reminiscent of alarm calls among primates. Standard CST provides an explanation for some forms of alarm calls directed at non-kin [*21*]. The hypertrophy of human language may have been, from the start, a consequence of the total unpredictability of danger in our species. Unexpectedness, defined as abnormal structure [*10, 27*], is the only signature of potential killing danger. This may contribute to explaining why information has replaced muscles in hominin social displays, and why humans can provide information about 'everything'.

# Annex

The emergence of stable communication strategies is studied both theoretically and through computer simulation. We will first provide details about the algorithm used to implement the Friendship Signaling Model; we then give a theoretical account showing that two evolutionary stable strategies (ESS) are expected to emerge; lastly, we compare these theoretical predictions with simulation results.

## The Friendship Signaling Model and its Implementation

### Social Offer and Social Benefit

Note: Unless stated otherwise, all quantities belong to the segment [0,1]. In figures, they are displayed in percentages, between 0 and 100.

A population of agents differing by their ability $q \in [0,1]$ to send relevant information interact and establish social links. Each agent emits a signal $s(q) = g(q) \, S(q)$, where $g(q) \in [0,1]$ is its investment in communication and $S(q) = b + (1–b) \, q$ is its communicative competence. Parameter $b \in [0,1]$ is introduced to represent the fact that worst individuals may still possess a bottom competence. At each step, a pair (A,B) of agents is selected and plays the following game. Agent A (with ability $q_A$) makes a social offer $s(q_A) \, r^i$, where $i$ is the rank offered by A to B in its friendship shortlist (with $0 \le r \le 1$). A starts by offering $i$=0, but it may then increase $i$ if B's return offer is smaller than what A's current friend at rank $i$ offered in a previous encounter. B follows the same rule. If A and B come to an agreement, they insert the partner at the $i^{\text{th}}$ place in their friendship list.

Agents play the game a number of times (typically 300 encounters for 100 agents) at each step. Then agents compute the benefit obtained from having friends. The following formula gives the social profit that an agent who could make friends with abilities $q_i$ gets:

$$P = (1 – \Pi(1 – K \, r^i \, S(q_i)))$$

The product is computed over all the agent's current friends. We could have used a variety of increasing functions of the friends' competence $S(q_i)$. The above formula is dictated by the protection scenario (see main paper). The positive term in the above formula represents the risk for the agent to get killed by surprise. The $i^{\text{th}}$ friend has a probability $K \, S(q_i)$ of spotting a potential risk at each moment, and this protection covers a portion $r^i$ of the day.

Agents endure a proportional cost for signaling. The overall benefit of an agent with ability $q$ is:

$$B = P – C \, g(q)$$

*Learning*

We designed a simple learning algorithm that optimizes $g(q)$. One agent learns at each step, after the tournament has been played. The learning agent adopts a new value for $g(q)$ that realizes a compromise between various values:

- $g(q')$ for neighboring abilities $q'$.

- the past value of $g(q)$ that provided the highest value of $B$ (memory span is typically limited to 10 past learning episodes).

- an additive perturbation of $g(q)$ of amplitude $L$. $L$ decreases until the agent reaches 'adulthood', where it reaches a bottom value $L_0$.

Agents 'die' when they reach a maximum age. They are replaced by another agent with same ability $q$ but a random value for $g(q)$. After a definite number of steps, the overall shape of function $g$ is supposedly reached. All new agents are then born adult, as a way to lower the temperature of the learning system.

The learning process is illustrated in the joined animated image 'AnimateSignal.gif', also available at www.dessalles.fr/Evolife/friendship/AnimateSignal.gif .

## Evolutionary Stable Signaling Strategies

*Competitive Signaling*

Here we show that negotiation about social offer $s(q)\, r^i$ may lead to a situation where competitive signaling and assortative bonding reinforce each other. Competitive (or honest) signaling means that $s(q)$ is an increasing function. By emitting $s(q_A)$, agent A attracts a friend B with ability $q_B = f_c(s(q_A))$, where $f_c$ is a non decreasing function. Suppose $q_B = q_A - \delta_1$ ($\delta_1 > 0$). Since B is acquainted with A, $f_c(s(q_B)) = q_A$. Since $s$ is increasing, we get: $f_c(s(q_A - \delta_1)) = q_A < f_c(s(q_A)) = q_A - \delta_1$, which is a contradiction. A similar reasoning with $\delta_1 < 0$ leads to the conclusion that $f_c(s(q_A)) = q_A$, which means that social bonds are assortative.

Conversely, if there is assortativeness, then $s$ evolves toward a definite increasing function. An individual with ability $q$ gets social profit $P(q) = (1 - \Pi(1 - K\, r^i\, S(q)))$. Learning maximizes the benefit:

$$B_c = P(q) - C\, s(q)/S(q)$$

If the individual sends the signal normally sent by agents with quality $q+dq$, both profit and cost vary:

$$dB = P(q+dq) - C\, s(q+dq)/S(q)$$

At the maximum, $dB/dq = 0$, which gives:

$$s'(q) = S(q)\, P'(q)/C$$

and finally:

$$s(q) = [ \; S(q)P(q) - (1-b) \int P(q) \; dq \; ] \; /C$$

In the absence of noise, any competitive situation in which $s(0) \neq 0$ is not an ESS, as lowest individuals (with $q \approx 0$) are bound to establish links with each other anyway. A mutant with $s(0) = 0$ would endure no cost for the same profit.

If the number $n$ of friends per agent is limited, we get the following polynomial functions (for $b = 0$):

If $n = 1$: $\qquad s(q) = K \; q^2 \; /(2C)$

If $n = 2$: $\qquad s(q) \; = \; (-2K^2 r \; q^3/3 + K \; q^2(1+r) \; /2) \; /C$

If $n = 3$: $\qquad s(q) \; = \; ( \; K(1+r+r^2) \; q^2/2 - 2K^2 r(1+r+r^2) \; q^3/3 + 3K^3 r^3 \; q^4/4 \; )/C$

Benefit in the competitive mode is:

$$B_c = (1-b) \int P(q) \; dq \; / \; S(q)$$

*Uniform signaling*

In the uniform mode, all individuals with ability above $\eta$ send the same signal $s_m$. Since they cannot distinguish each other, they get acquainted with a random member of the $[\eta, 1]$ 'elite club'. On average, their profit is $P(\tau)$, where $\tau = (1+\eta)/2$. Their benefit depends on their ability $q$:

$$B_u = P(\tau) - C \; s_m / \; S(q)$$

Signaling above $s_m$ would increase cost without providing any profit. Signaling below $s_m$ would spare a tiny share of the cost but profit would drop from $P(\tau)$ to $P(\eta)$. Using this reasoning, $s_m$ is thus an ESS, whatever its value. Simulations, however, always converge to a definite value of $s_m$. To explain the phenomenon, we must take into account the inevitable uncertainty about $s_m$ introduced by learning.

The variability of the signal emitted by agents with ability $q$ depends on the amplitude of learning. We may consider that it is $\alpha L_0 S(q)$, where $L_0$ is the maximum variation of $g(q)$ in one learning step, starting from the best previously encountered value. An agent in the $[\eta, 1]$ range emits $(s_m + \rho \alpha L_0 \; S(q))$, where $\rho \in [-1, 1]$. Its probability of getting acquainted with another agent of the elite club, and thus of getting social profit $P(\tau)$, varies between 0 for $\rho = -1$ and 1 for $\rho = 1$. A linear approximation gives the following expression for the benefit:

$$B_u(\rho) = (1+\rho) \; P(\tau)/2 + (1-\rho) \; P(\eta)/2 - C \; (s_m + \rho \alpha L_0 \; S(q))/ \; S(q)$$

We get:

$$\mathrm{d}B_u/\mathrm{d}\rho = P(\tau)/2 - P(\eta)/2 - C \alpha L_0$$

$\mathrm{d}B_u/\mathrm{d}\rho$ must be zero, otherwise $s_m$ would not be stable.

$$P(\tau) - P(\eta) = 2C \alpha L_0$$

This relation defines $\eta$. The threshold $\eta$ corresponds to the limit between the competitive mode and the uniform mode. We can write that $B_u$ and $B_c$ are equal in $\eta$.

$$(P(\tau) + P(\eta))/2 - C\ s_m/\ S(\eta) = (1-b) \int_0^\eta P(q)\ dq\ /\ S(\eta)$$

This relation defines $s_m$.

*Competitive–Uniform transition*

As soon as there is a discontinuity in $\eta$ between $s_m$ and the competitive signal $s_c$, we would expect a sharp transition. Such a transition is not to be observed, however, as individuals with ability in the range $[\theta, \eta]$ adopt yet another ESS. This intermediary ESS consists in emitting signal $(s_m - \sigma)$ on average. The benefit on this new ESS is:

$$B_u(\rho) = (1+\rho)\ P(\eta)/2 + (1-\rho)\ P((\theta+\eta)/2)/2 - C\ (s_m - \sigma + \rho\alpha L_0\ S(q))/\ S(q)$$

By making $dB_u/d\rho = 0$, we get:

$$(P(\eta) - P((\theta+\eta)/2)) = 2C\ \alpha L_0$$

This gives a minimum value for $\theta$, given the constraint that $B_u > B_c$ in $\theta$.

The same reasoning can be iterated for various couples $(\sigma_i, \theta_i)$, which explains the smooth transition that can be observed between the competitive and the uniform mode.

## Observations

In Figure 3, parameters are: $n = 3$, $r = 0.6$, $K = 1$, $b = 0$, $L_0 = 0.05$. Each point of the curves is the average of 30 experiments at least. The only parameter of the model is $\alpha$, which has been set to 1.2. Transitions for the model are shown in $\theta$.

Figure A1 shows, for the same parameters, the simple case in which individuals can have only one friend. The parabolic shape of competitive signals can be clearly seen.

Figure A2 shows the observed and computed values of the uniform signal $s_m$ depending on $L_0$ (A) and $r$ (B). The model's predictions are nearly perfect for these values of the cost coefficient.

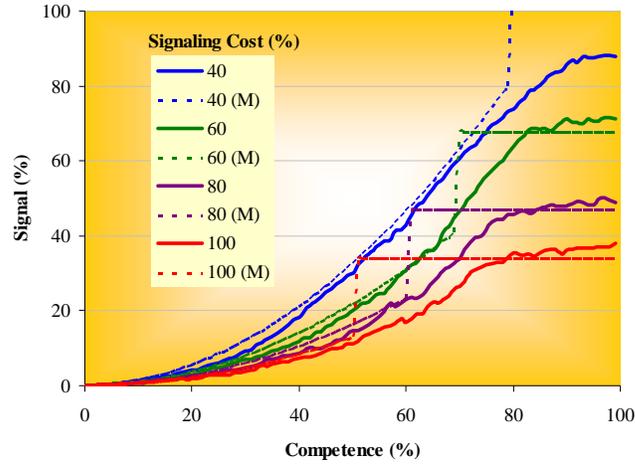

**Figure A1.** Average signals for various values of *C*.
Dashed lines show model predictions. The number of friends per individual is limited to 1.

Note that the model tends to overestimate $s_m$ when $C \leq 0.5$, or even does not predict its existence, as can be seen on Figure A1 for $C = 0.4$ and $n = 1$. This is because $g(q)$ saturates ($g(q) = 1$) for intermediary values of $q$, what artificially limits competition level for higher values of $q$.

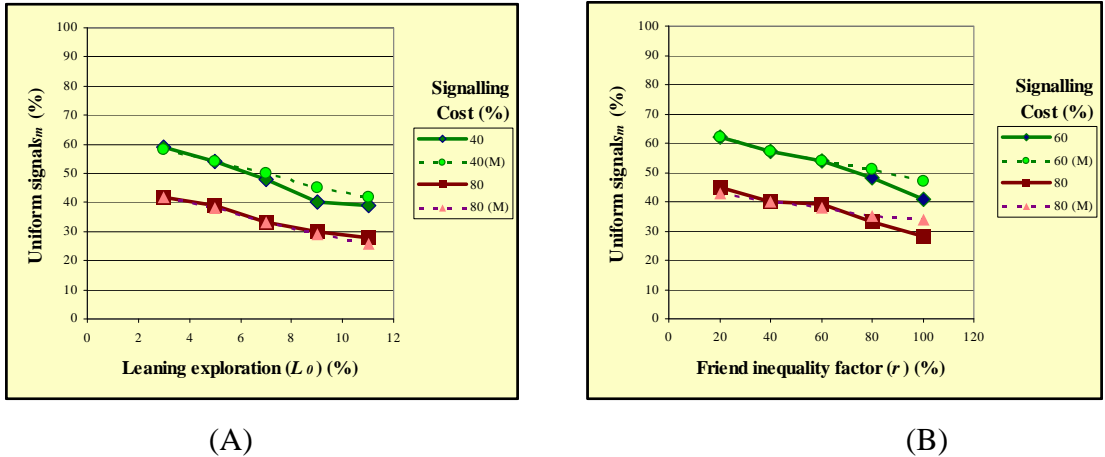

(A)                                                                                    (B)

**Figure A2.** Uniform signal values.
The figure shows $s_m$ depending on the amplitude of learning exploration (A) and on the friend inequality factor (B). Dashed lines show model predictions. The number of friends per individual is limited to 2. ($K = 1$; $b = 0$; $r = 0.6$ in (A); $L_0 = 0.05$ in (B)).

Figure A3 shows that when the maximum number of friends per individual is not limited or has a significant value (here $n = 10$), signaling vanishes if all friends are given the same amount of time whatever their rank ($r$ close to 1).

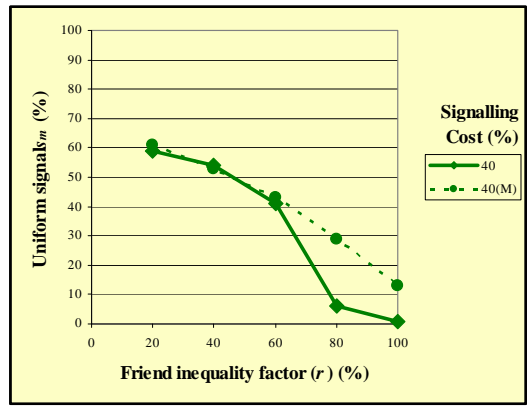

**Figure A3.** Uniform signals depending on friend inequality factor ($r$).
Individuals may make up to 10 friends ($n = 10$). The dashed line represents the model predictions. ($K = 1$; $b = 0$; $L_0 = 0.05$).

## *Simulation program*

The simulation program is available at [www.dessalles.fr/Evolife/friendship](http://www.dessalles.fr/Evolife/friendship)